\documentclass[a4paper,
               keeplastbox,   
               ]{jacow}
\usepackage{pdfpages,multirow,ragged2e} %

\makeatletter%
	\ifboolexpr{bool{xetex}}
	 {\renewcommand{\Gin@extensions}{.pdf,%
	                    .png,.jpg,.bmp,.pict,.tif,.psd,.mac,.sga,.tga,.gif,%
	                    .eps,.ps,%
	                    }}{}
\makeatother

\ifboolexpr{bool{xetex} or bool{luatex}} 
 {}                                      
 {\usepackage[utf8]{inputenc}}           

\usepackage[USenglish]{babel}

\ifboolexpr{bool{jacowbiblatex}}%
 {%
  \addbibresource{jacow-test.bib}
  \addbibresource{biblatex-examples.bib}
 }{}
\begin{document}

\title{Analog Cavity Emulators to Support LLRF Development\thanks{This work was supported by the LCLS-II and ALS-U Projects and the U.S. Department of Energy, Contract DE-AC02-76SF00515}}

\author{S.D. Murthy\thanks{sdmurthy@lbl.gov}, L. Doolittle,  LBNL, Berkeley, CA 94720, USA\\
A. Benwell, SLAC, Stanford, CA 94309, USA }

\maketitle

\begin{abstract}
The goal of a LLRF system is to control an actual RF cavity with beam. While digital simulations have a place, having an analog circuit to stand in for the cavity can be tremendously helpful in validating hardware+firmware+software under development. A wide range of cavity emulators have been developed in collaboration with SLAC, and LBNL. Cavity emulators are typically based on quartz crystals and frequency conversion hardware. The choice of crystal frequency and coupling mechanism depends in part on the bandwidth and coupling of the cavity it’s intended to emulate. Examples of bandwidth range from 800~Hz (SLAC) as a stand-in for a SRF cavity, to 31~kHz (LBNL) for a room-temperature accumulator-ring cavity. An external LO is used to tune the emulated cavity frequency. The coupling properties are also of interest if the scope includes emulating reverse power waveforms. LLRF system checks such as closed-loop bandwidth, and determining cavity detuning can be performed interactively and as part of a Continuous Integration (CI) process. This paper describes the design, implementation, and performance of the cavity emulators.
\end{abstract}

\section{INTRODUCTION}
At most of the accelerator facilities, the real RF cavities will still be under construction when the LLRF system is being developed. Even when the cavities become available, setting up the whole high-power system (e.g., cryomodule) just to validate the LLRF system under development is difficult to justify. People also worry about damaging the cavities with an unproven controller.
Digital simulations are a great tool at the beginning of the development process to understand and test the basic functionality of the firmware \cite{IEEE}. In light sources, requirements flow from the quality of the X-ray beams to the electron beam quality and then to requirements on the stability of the accelerating cavity fields. Examples for the stability requirement range from $0.01^{\circ}$ in phase and $0.01\%$ in amplitude for SLAC's Linear Coherent Light Source Linac (LCLS-II) system \cite{lclsii} and $0.1^{\circ}$ in phase and $0.1\%$ in amplitude for LBNL's Advanced Light Source Upgrade (ALS-U) system. Careful engineering of the entire LLRF system is required to attain this stability requirement.

To support the development process, cavity emulators have been developed at various accelerator labs over the years to validate the LLRF system before controlling a real RF cavity. A collaboration between SLAC and LBNL has resulted in the design of cavity emulators for SRF cavities with 800~Hz bandwidth for LCLS-II and room temperature accumulator-ring cavities with 31~kHz bandwidth.

\section{Cavity Emulator Design}
\begin{figure}[!htb]
   \centering
   \includegraphics*[width=1.15\columnwidth]{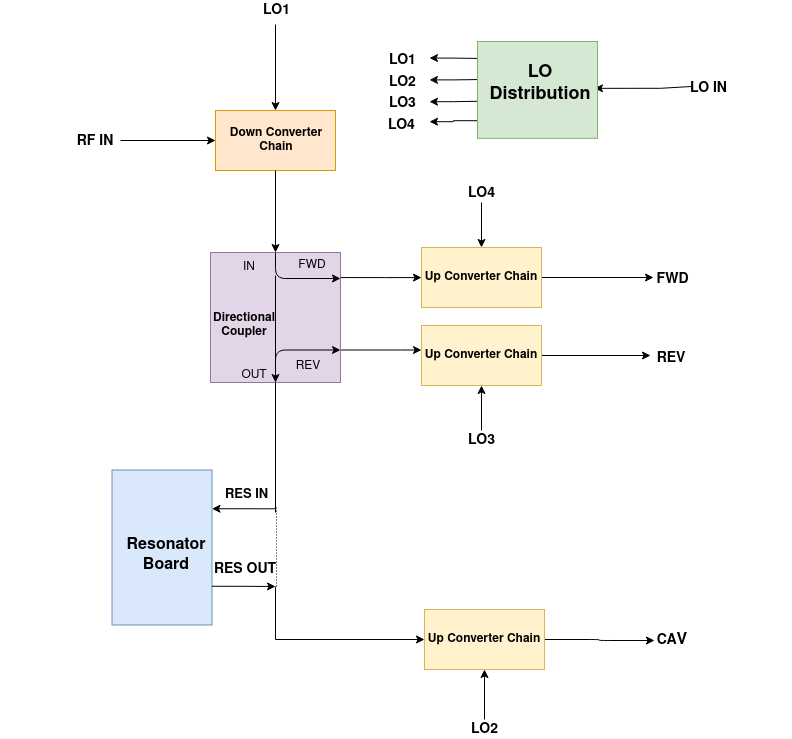}
   \caption{Cavity emulator components.}
   \label{fig:block_diagram}
\end{figure}

An analog cavity emulator consists of two components: a crystal resonator, and frequency conversion chains as shown in Fig.~\ref{fig:block_diagram}. They can either be constructed on Printed Circuit Boards (PCBs), or with componentized parts, or a combination of both.

\subsection{Resonator hardware}
The resonator consists of a commercial quartz crystal, often with additional passive components.
A quartz crystal is normally modeled as a series RLC circuit with a parallel capacitor. The former represents the mechanical vibrations of the crystal, and the latter represents the electrodes attached to the crystal\cite{wiki}.
A crystal has two resonance modes, series and parallel. The frequency at which the series inductance and series capacitance combination exhibit a phase angle of zero is defined as series resonance frequency ($f_s$). A parallel resonance ($f_p$) is created when series inductance and capacitance interact with parallel capacitance.

All crystal parameters, such as Equivalent Series Resistance (ESR), Shunt Capacitance ($C_0$), Motion Inductance ($L_m$), Motion Capacitance ($C_m$), and Quality factor ($Q_L$) need to be characterized using the measured transfer function ($S_{21}$) to achieve the desired bandwidth and minimize broadband coupling \cite{IEEE1}.
\begin{equation}\label{eq:label1}
    ESR = 2 \cdot Z_0 \cdot (1/|S_{21}|-1)
\end{equation}

\begin{equation}\label{eq:label2}
    R_{\rm eff} = 2 \cdot Z_0 + ESR
\end{equation}

\begin{equation}\label{eq:label3}
    \Delta f = f_h - f_l
\end{equation}

\begin{equation}\label{eq:label4}
    C_m = \frac{\Delta f}{2 \cdot \pi \cdot f_s^{2} \cdot R_{\rm eff}}
\end{equation}

\begin{equation}\label{eq:label5}
    L_m = \frac{R_{\rm eff}}{2 \cdot \pi \cdot \Delta f}
\end{equation}

\begin{equation}\label{eq:label6}
    Q_L = \frac{2 \cdot \pi \cdot f_s \cdot L_m}{ESR}
\end{equation}

Here $Z_0$ is the transmission line impedence, $R_{\rm eff}$ is the effective resistance across the crystal, $f_h$ and $f_l$ are the frequencies at $+45^{\circ}$ and $-45^{\circ}$ $S_{21}$ phase respectively.

The complex $S_{21}$ of the crystal is given by equation \eqref{eq:label7}. When $\beta_0$, $\beta_1$ coefficients are zero, it becomes a pure complex Lorentzian function equation \eqref{eq:label8}. If there were no parasitics, and the network analyzer calibration was perfect (including correcting for the test set), there would be no additional phase shift or offset required.
$\beta_0+j\beta_1$ are broadband coupling coefficients, $\beta_2+j\beta_3$ are scaling and rotation coefficients due to Lorentzian \cite{IEEE2}.
A wide range of commercially available crystals have been characterized, using NumPy and Jupyter-Notebook to curve-fit the acquired $S_{21}$ data.
\begin{equation}\label{eq:label7}
S_{21} = \beta_0 + j\beta_1+ (\beta_2 + j\beta_3 )\cdot a
\end{equation}

\begin{equation}\label{eq:label8}
 a =\frac{1}{1+2j \cdot Q_L \cdot (\frac {f}{f_s} -1)}
\end{equation}

One way to reduce the broadband coupling is to add a compensating inductance in parallel with the crystal.  That inductance cancels the crystal's parasitic shunt capacitance.  For trimming purposes, one can either use a variable inductor (SLAC), or use a fixed inductor with an additional variable capacitor (LBNL) .
A series resistance can also be added to adjust the circuit's bandwidth. If this load resistance increases, the signal strength decreases and the bandwidth increases. But some loss in the signal strength is acceptable, because the upconversion chains need to be designed to handle forward and reverse signals attenuated by the directional coupler.

\begin{figure}[!htb]
   \centering
   \includegraphics*[width=1.0\columnwidth]{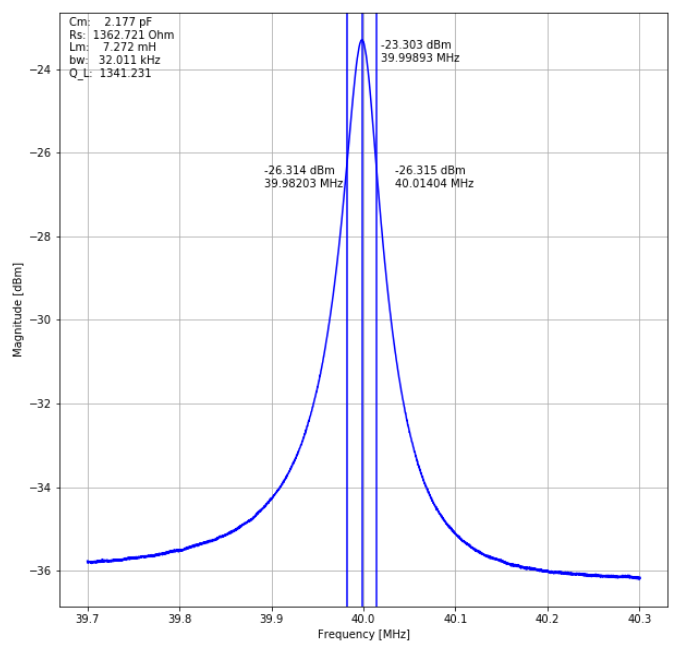}
   \caption{Crystal resonance frequency (LBNL) (span = 600~kHz).}
   \label{fig:resonator_output}
\end{figure}

\begin{figure}[!htb]
   \centering
   \includegraphics*[width=1.0\columnwidth]{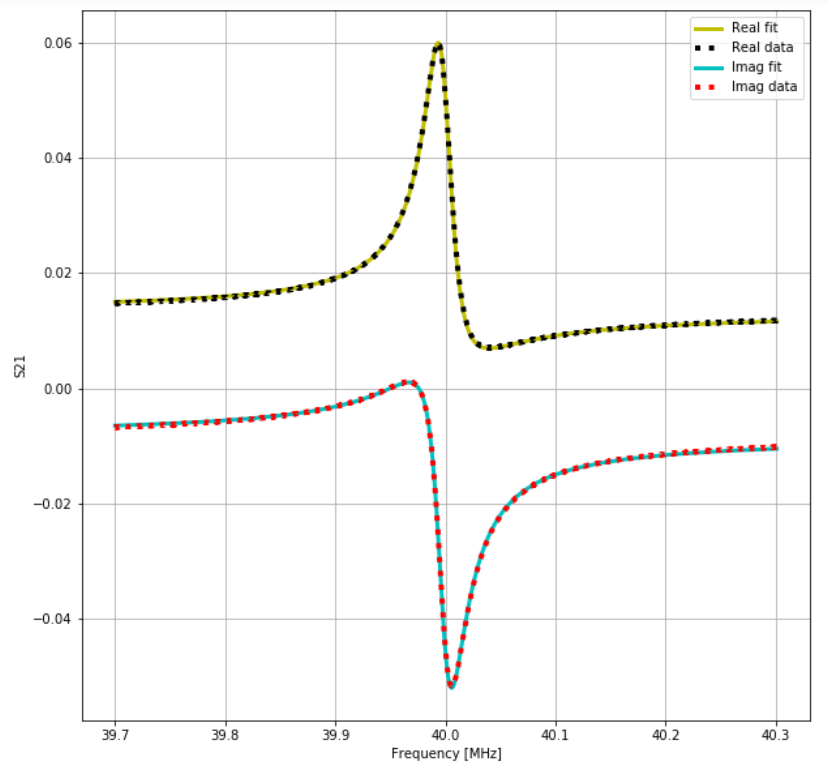}
   \caption{Complex frequency response with curve fitting.}
   \label{fig:resonator_curvefit}
\end{figure}

Fig.~\ref{fig:resonator_output} shows the frequency response measurement of the series resonance for the
LBNL cavity emulator. The measured loaded $Q_L$ is 1341 (which corresponds to a full-bandwidth of about
31~kHz), and the insertion loss is 23\thinspace dB. The complex frequency response $S_{21}$ of the curve-fitted data is Fig.~\ref{fig:resonator_curvefit}. Similar frequency response curves can be derived for any available commercial crystals.

\subsection{Frequency conversion hardware}
The frequency conversion hardware consists of a single down conversion chain and three up conversion chains. The down converter is used to convert the incoming RF cavity drive signal to the crystal resonance frequency. The emulated cavity output is then up converted to RF. Forward and reverse signals are also be produced with a directional coupler at IF, and then up converted. Special care is taken to select a well-defined bandpass filter to remove any spurs/leakage after the up-conversion chains. An external signal source is used as a LO to tune the emulated system's resonance frequency.

\section{Performance}
\begin{figure}[!htb]
   \centering
   \includegraphics*[width=1.0\columnwidth]{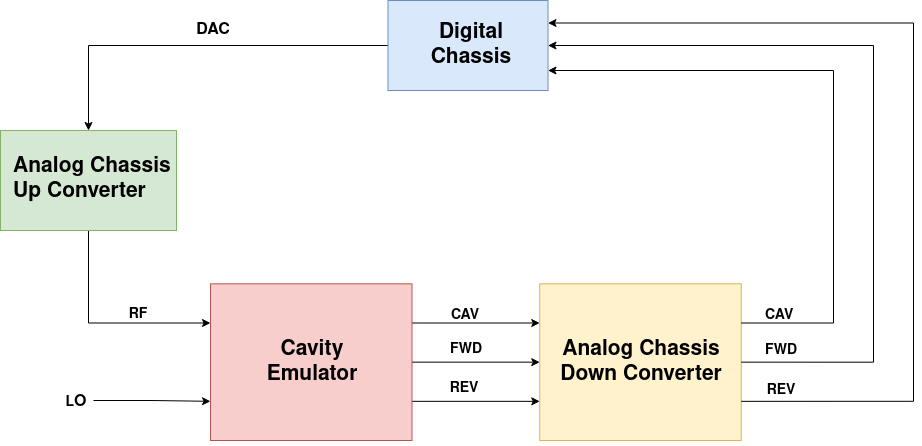}
   \caption{Cavity emulator integrated with LLRF system.}
   \label{fig:llrf}
\end{figure}

\begin{figure}[!htb]
   \centering
   \includegraphics*[width=1\columnwidth]{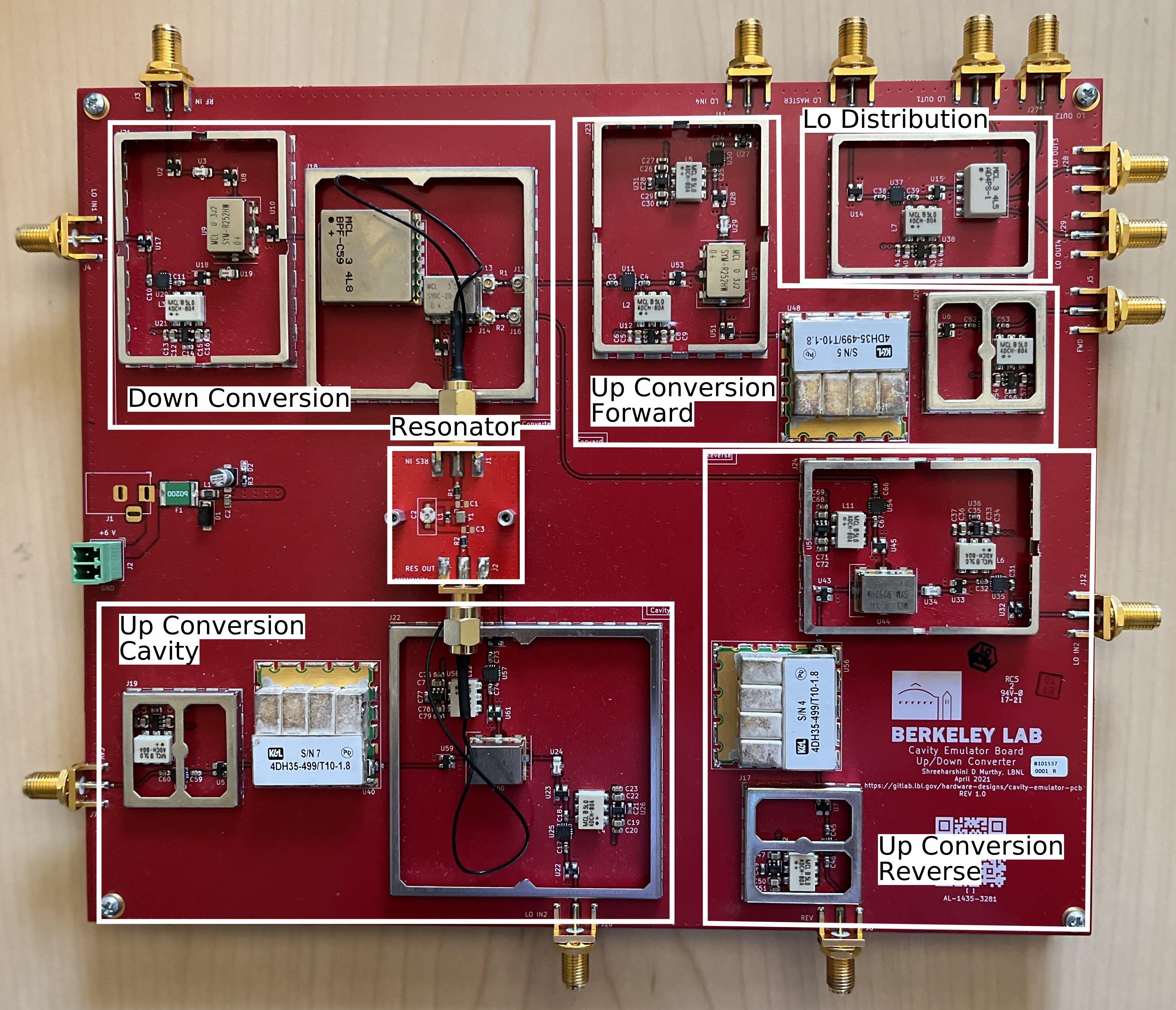}
   \caption{Cavity emulator PCB at LBNL.}
   \label{fig:pict}
\end{figure}

\begin{figure}[!htb]
   \centering
   \includegraphics*[width=1\columnwidth]{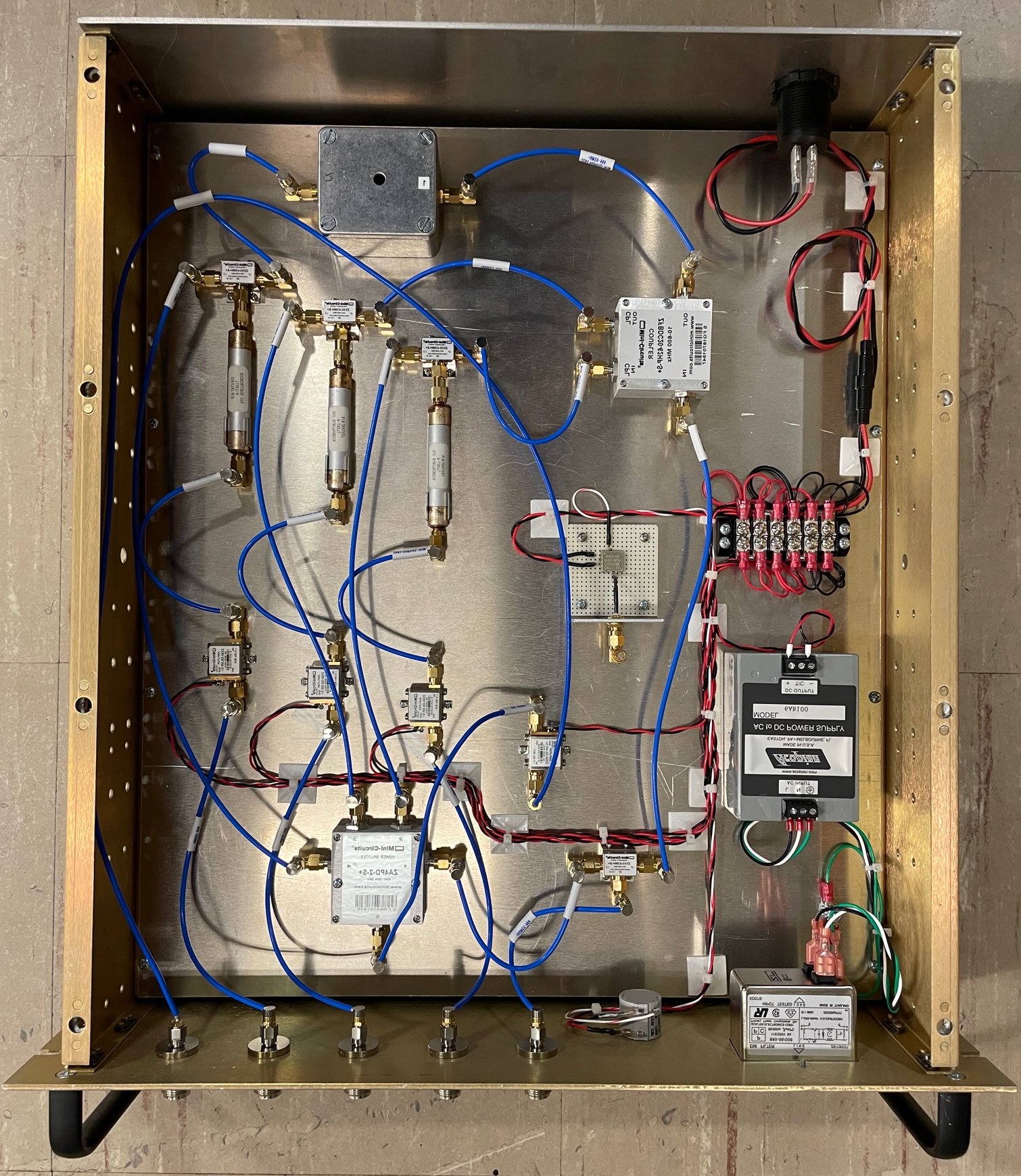}
   \caption{Componentized cavity emulator at SLAC. }
   \label{fig:pict2}
\end{figure}

The cavity emulators designed to emulate SLAC's SRF cavity and LBNL's accumulator-ring were integrated with the LLRF system as shown in Fig.~\ref{fig:llrf}, along with picture of the PCB (LBNL) is shown in Fig.~\ref{fig:pict} and the picture of the componentized cavity emulator (SLAC) in Fig.~\ref{fig:pict2}.

Several emulators were deployed for Continuous Integration (CI) purposes. A server runs automated tests to ensure that any changes to hardware/firmware/software don't break the existing codebase. By extension, the codebase is always checked on the test stand and can be deployed to the rest of the LLRF system without any issues.
These automatic tests include bringing up the cavity emulator and closing the RF feedback loop. The closed-loop bandwidth of the controllers is also measured along with the total group delay. The in-loop stability analysis for both the SLAC and LBNL system also meets the requirement. The cavity detuning function was also characterized by changing the LO frequency.

Complete characterization of a LLRF system cannot be done using the cavity emulator; this includes the out-of-loop stability analysis. The Signal-to-Noise Ratio (SNR) of a cavity emulator is worse than real cavities, and there might be spurs caused by LO leakage.

\section{CONCLUSION}
Cavity emulators were designed and built in collaboration with SLAC, and LBNL. They continue to offer a good benchtop evaluation for testing LLRF systems including the hardware+firmware+software and train system engineers to understand the loop parameters better. Emulation of different RF cavities can also be achieved with the existing system by using a different crystal, coupling mechanism, and up conversion bandpass filters.

%
%
\ifboolexpr{bool{jacowbiblatex}}%
	{\printbibliography}%
	{%

	
} 
%
%


\end{document}